\documentclass[12pt]{article}

\usepackage{scicite}

\usepackage{times}

\usepackage{graphicx}

\newenvironment{sciabstract}{%
\begin{quote} \bf}
{\end{quote}}

\newcounter{lastnote}
\newenvironment{scilastnote}{%
\setcounter{lastnote}{\value{enumiv}}%
\addtocounter{lastnote}{+1}%
\begin{list}%
{\arabic{lastnote}.}
{\setlength{\leftmargin}{.22in}}
{\setlength{\labelsep}{.5em}}}
{\end{list}}

\title{Primordial Gravitational Waves and Cosmology}

\author
{Lawrence Krauss,$^{1\ast}$ Scott Dodelson,$^{2,3,4}$ Stephan Meyer $^{3,4,5,6}$\\
\\
\normalsize{$^{1}$School of Earth and Space Exploration and Department of Physics,
Arizona State University,}\\
\normalsize{PO Box 871404, Tempe, AZ 85287\vspace{0.1in}}\\
\normalsize{$^{2}$Center for Particle Astrophysics, Fermi National Laboratory}\\
\normalsize{PO Box 500, Batavia, IL 60510\vspace{0.1in}}\\
\normalsize{$^{3}$Department of Astronomy and Astrophysics}\\
\normalsize{$^{4}$Kavli Institute of Cosmological Physics}\\
\normalsize{$^{5}$Department of Physics}\\
\normalsize{$^{6}$Enrico Fermi Institute}\\
\normalsize{University of Chicago, 5640 S. Ellis Ave, Chicago, IL 60637\vspace{0.1in}}\\
\normalsize{$^\ast$To whom correspondence should be addressed; E-mail:  krauss@asu.edu.}
}

\date{}

\begin{document} 


\baselineskip24pt


\maketitle


\begin{sciabstract}
The observation of primordial gravitational waves could provide a new and unique window on the earliest moments in the history of the universe, and on possible new physics at energies many orders of magnitude beyond those accessible at particle accelerators. Such waves might be detectable soon in current or planned satellite experiments that will probe for  characteristic imprints in the polarization of the cosmic microwave background (CMB), or later with direct space-based interferometers. A positive detection could provide definitive evidence for Inflation in the early universe, and would constrain new physics from the Grand Unification scale to the Planck scale.
\end{sciabstract}


\section*{Introduction}

Observations made over the past decade have led to a standard model of  cosmology: a flat universe dominated by unknown forms of dark energy and dark matter.   However, while all available cosmological data are consistent with this model, the origin of neither the dark energy nor the dark matter is known.  The resolution of these mysteries may require us to probe back to the earliest moments of the big bang expansion, but before a period of about 380,000 years, the universe was both hot and opaque to electromagnetic radiation.  Thus to probe back to earlier times we need to search for other observables outside of the electromagnetic spectrum.

Gravitational waves, predicted in General Relativity, interact very weakly with matter, and therefore the universe has been transparent to them since shortly after the big bang.  As a result, they provide a potentially new probe of early universe cosmology--that is, if we can detect them.  Here we review the most likely early universe sources of a detectable gravitational wave background and also explore the most likely detection method, via a measurement of polarization in the Cosmic Microwave Background (CMB) radiation--remnant thermal radiation from the Big Bang that has propagated unimpeded to our detectors since the time that the universe first became transparent to electromagnetic radiation.

\section{Primordial Gravitational Waves}

Just as in Electromagnetism moving charges can generate electromagnetic waves, in General Relativity moving masses can generate gravitational waves.  Because of the weakness of gravity however, in order to generate waves on a scale that might actually be detectable, astronomical amounts of matter must be moved around.  Direct detectors such as the LIGO detector \cite{LIGO} are being designed to measure the local space-time distortion generated by gravitational waves from astrophysical objects such as colliding neutron stars, or black holes. 

There is another cosmological source of gravitational waves, however, which can arise from the dynamics of our expanding universe at very early times.  Such a background will involve waves today whose wavelength will extend all the way up to our present cosmological horizon (the distance out to which we can currently observe in principle), and which are likely to be well beyond the reach of any direct detectors for the foreseeable future.

Primordial gravitational waves also leave indirect signatures that might show up in CMB maps.  Primordial mass and energy density fluctuations grow to produce cosmological structures, and also create observable CMB temperature anisotropies.  Primordial gravitational waves, on the other hand, produce not only temperature anisotropies but also a distinctive signal that can be detected with very sensitive measurements of the polarization of the CMB. While temperature anisotropies have been observed, current experiments have only been able to put upper limits on polarization anisotropies that might result from a gravitational wave background.

There are reasons to believe, however, that such a background exists and should be observable in the next generation of CMB polarization experiments.   This is because the observed
pattern of temperature anisotropies, combined with that of inhomogeneities in matter on large scales, and measurements of the total energy density in the universe are in striking agreement with the simplest predictions of so-called Inflationary models \cite{Guth:1980zm, Linde:1981mu, Albrecht:1982wi, Linde:1983gd}.  

Inflation suggests that early in its history, the universe underwent a brief epoch of accelerated expansion during which small-scale fluctuations were stretched to superhorizon scales. The theory posits that quantum mechanical fluctuations in the fields that describe the matter content at early times seeded the structure in the universe, leading to two important signatures: (a) Primordial anisotropies imprinted in the CMB at a level of a few parts in a hundred thousand (fig. 1), and (b)  
An inhomogeneous matter distribution in the form of a cosmic web as traced for example by the location of galaxies in the universe (see fig. 2). 

The inflationary paradigm is remarkably consistent with the detailed statistical features of both of these observed signatures, which can arise from scalar density perturbations generated during an early inflationary phase.   But scalar density perturbations are not the only perturbations that are generated.  
Gravitational waves are also produced (see SOM for details).  

When some background field gets stuck in a metastable state, energy gets stored in space, acting like a cosmological constant, producing a negative pressure and a positive energy density.  Solving Einstein's equations in such a background results in an accelerated expansion which quickly becomes exponential.  The appropriate metric for such a universe is the de Sitter metric.  

In a de Sitter background all massless or light quantum fields will fluctuate with a magnitude which is proportional to the only dimensional parameter governing the expansion, the stored energy density, $\rho$, which by Einstein's equations is related to the square of the Hubble expansion parameter, $H$ at the time.  This implies that a fluctuating background gravitational waves, which in the quantum theory correspond to massless quantum fields, will also be generated.  As inflation stretches distances, their wavelength soon becomes much larger than the Hubble radius --they are driven outside the horizon -- after which their amplitude remains fixed since no causal process can act over distances larger than the horizon.

After inflation ends, these modes return inside the horizon as a stochastic background of gravitational waves with a power that is proportional to $H^2$ \cite{rubakov, starob2, krauwhite, whitepap}.  Clearly, the larger the energy scale at which inflation occurs, the larger will be the residual gravitational wave signal.   This signal is thus determined solely by the energy scale at which inflation may
have occurred.

In the context of the CMB, as we shall describe, such modes can leave an imprint in the polarization of radiation.  One can then compare the predicted polarization power spectrum to the power spectrum for direct observed temperature anisotropies in the CMB  (see SOM1)

The quantity $r$, defined as the ratio of these two power spectra \cite{whitepap}, plays a definitive role in parametrizing the detectability of gravitational waves in CMB experiments.  The next generation of detectors is being designed with the intention of reaching a sensitivity to $r \ge .01$.   For the simplest models involving a single field, once $r$ is measured, the energy density during inflation will also be determined; the two are related via $ V^{1/4} = 1.06 \times 10^{16}\, {\rm GeV}\, { \left ( {r / 0.01} \right ) }.$
Current limits of $ r <  0.3$ already have severely constrained or ruled out a variety of single field inflationary potentials and give hope that increasing the experimental sensitivity to $r$ could yield a positive non-zero signature for gravitational waves in the near future. 

At the same time, however, there is an independent argument that suggests one should be cautious before counting on such a possibility.   Inflation requires an expansion factor of at least $e^{60}$ in order to resolve all the cosmological puzzles that it was designed to resolve.  One can ask what the total field excursion is during this period, and the larger the scale of inflation, the larger the magnitude of the inflaton field at the end of inflation.  One finds, in fact, again for single field inflation \cite{whitepap}:
$
{ \Delta \phi/ M_{pl}} \ge 1.06 \times {\left ( r/ .01 \right )}^{1/2}
$
where $M_{pl}$ is the Planck mass.  Field excursions larger than the Planck scale lead to territory where Planck Scale effects may be significant, producing possibly large undetermined corrections to simple quantum field theoretic estimates.  In fact, in many string theory models the value of the inflating field is associated with an excursion in an extra dimension, which is itself restricted to be Planck length in size so that it is  impossible for the field to take values larger than the Planck scale, and one would expect, therefore that $r <.01$.   More recently it has been realized that string theory might allow for cases where $r$ can exceed 0.01  \cite{silv} so that distinguishing between small and large $r$ would help define the symmetry structure of the theory.  Thus  $r=0.01$ represents an important threshold.  

Because inflation produces gravitational waves on all scales, one might hope that other probes of gravitational waves with smaller wavelengths and higher frequencies might be useful. But once inside the horizon, gravitational waves redshift like radiation.  During matter domination, the contribution to the total energy density in such modes falls by a factor identical to that of photons, so that their contribution to the energy density today will be suppressed.  

Finally,  Inflation is not the only mechanism that might produce a stochastic gravitational wave background.  For example, a phase transition in the early universe could result in a background whose spectrum inside the horizon is remarkably similar to that from inflation \cite{krauss1}. Detailed calculations suggest that the signal from such a transition can be enhanced  \cite{jkm, kjmd, durrer}, so that even transitions somewhat below the scale of inflation could produce a comparable signal.  Various different possible ways of distinguishing this signal have been suggested including looking at the real space correlations of polarization \cite{baum2}, exploring the three point correlation function for polarization\cite{lim}, and comparing a possible CMB gravitational wave signal with a future signal in direct detectors  \cite{kjmd}.

\section*{What can we learn from CMB Polarization?}

The greatest sensitivity to a primordial gravitational wave background comes from the detailed pattern of polarization in the CMB~\cite{kam,zal,huwhite}.  Thomson scattering of an anisotropic radiation background off of free electrons before the time of electron-proton recombination (when the universe was 380,000 years old) produced a radiation field polarized at the 10\% level. 
Because polarization is described at every position by an amplitude and an angle of orientation, the polarization field on the sky can be decomposed into two modes, a curl-free $E$-mode and a divergence-less $B$-mode, (Figure~3). The $B$-mode pattern cannot be produced by scalar perturbations, thus its detection would be a signature of primordial gravitational waves.

While the $E$-modes are predominantly generated from scalar perturbations, they are nonetheless important to those who study inflation: they have been detected, and the observed pattern of $E$-modes is in perfect agreement with the expectations from inflation. Inflation predicts that the temporal phases of all perturbations were set early in the history of the universe. This primordial coherence leaves its mark on the anisotropy pattern~\cite{dod}. Because polarization is produced by Compton scattering of anisotropic radiation off electrons, the spectrum of temperature anisotropies is predicted to be closely related to the $E$-mode spectrum. The $E$-modes reach their maximum and minimum amplitudes at precisely the same times as do the higher moments (dipole, quadrupole) of the radiation field. In contrast, the temperature pattern we observe is the monopole of the radiation field at recombination. The continuity equation dictates that the monopole and dipole are out of phase with one another, so the $E$-mode of polarization is $90^\circ$ out of phase with the acoustic pattern in the temperature spectrum (Fig. 4).  This phase lag  has been observed even on large scales of order a degree that were not in causal contact when the universe was 380,000 years old. The phases were set up at very early times, long before recombination, just as predicted by inflation.

The $B$-modes are the next frontier.  While the amplitude of the signal is unknown, the shape of the spectrum is robustly predicted by theory (Fig. 4). To understand the shape, recall that the polarization signal is generated only during those times when electrons are not trapped in neutral hydrogen, but are free to scatter off of radiation. However, if scattering is very efficient the resulting polarization signal will be very small. Imagine a
free electron in the early universe. Scattering off of this electron will produce polarization only if the observed incoming radiation field is anisotropic. If scattering is very efficient, though, the mean free path of photons will be very small and the last scattering surface that the electron observes will be very nearby. Thus all incoming radiation will share the same temperature and there will be no anisotropy from which to generate polarization. Polarization then requires scattering, but not rapid scattering. This set of requirements limits the epochs during which polarization is generated to two: (i) when electrons and protons are in the process of recombination at $t=380,000$ years old and (ii) much later when electrons are liberated from neutral hydrogen but the universe is so dilute that scattering is rare. The signals generated at these two times are imprinted onto the characteristic scales associated with each: the horizons $d_H(z=1100)$ and $d_H(z\simeq10)$. We observe these at angular scales equal to $\theta=d_H/d_A(z)$ where $d_A(z)$ is the angular diameter distance to redshift $z$. Hence, the predicted signal with bumps at angular scales of  $\sim 2$ and $\sim 50$ degrees. 

One complication which is also apparent in Fig. 4 is the effect of gravitational lensing~\cite{lens,lewis}. Scalar perturbations produce only $E$-modes at recombination, but the radiation from this last scattering surface travels through a clumpy universe before it reaches us. The ensuing gravitational lensing generates $B$-modes from the primordial $E$-field. These become important on smaller scales than those on which the $B$ humps appear, but they limit our ability for detect the higher $B$-harmonics. Although techniques have been proposed to clean this $E$-leakage, the most promising approach appears to be to measure the two peaks at $\sim 2$ and $\sim 50$ degrees.

\section*{Experimental Efforts, Present and Future}

The current generation of direct gravitational wave detectors such as the LIGO detector \cite{LIGO} do not have sufficient sensitivity to probe for possible primordial gravitational waves.  Direct detection will require more sensitive space-based interferometers, with a further improvement of at last $3-4$ orders of magnitude in sensitivity at the very least being required.  Because of the technological challenges and cost, such detectors are likely to take at least one or two decades to develop, and even then the likelihood of achieving the necessary sensitivity is not yet clear.  

Rapid progress in the measurement of CMB temperature and polarization fluctuation spectra may allow the detection of primordial gravitational waves within the decade (SOM2).  The experiments are carried out from ground-based observatories, high-altitude balloons and satellites in space.  Satellites operate for years making sensitive, all-sky maps which are not contaminated by atmospheric emission and can therefore measure large angular scale fluctuations. Satellite telescopes are by necessity small and therefore have limited resolution, angles above 0.2 degrees.  As with satellites, balloon-borne experiments are free of most of the interfering emission from the atmosphere and have the additional advantage of a relatively short development cycle. New, sensitive, high-pixel count focal planes currently give balloon experiments the highest instantaneous sensitivity but flights are limited to a couple of weeks at the maximum. Ground-based experiments on the other hand, can have large diameter telescopes making possible high angular resolution. The telescopes can operate for many years and have readily accessible instrumentation. These experiments have evolved from a first detection of temperature anisotropy in 1992, to the detection of CMB polarization in 2002 to sensitive probes of B-modes today, a gain in sensitivity of over $10^5$ in just 17 years.

Three elements combine to specify the overall power of experiments to probe the CMB inflation signal. The first is sensitivity. The CMB polarization signals are exceedingly weak, a few millionths of a degree but detectors must cope with the thermal emission of the telescope and environment. Only space instruments can be cold enough to avoid local emission. State of the art detectors today are limited by the fundamental photon noise and so the sensitivity of individual detectors cannot be improved much. However, large focal plane instruments with thousands of detectors, in effect, make many simultaneous measurements. Most of the improvement in instrument sensitivity over the past decade is the result of ever increasing detector count and thus instrument complexity. Plans for ground-based and satellite instruments with tens of thousands of detectors are underway and these will have the sensitivity needed to measure the two bumps in the $B$-mode polarization in Fig. 4.

The second element controlling the power of experiments is their ability to differentiate CMB structure from structure in foreground emission, particularly diffuse emission from our own Galaxy. Unfortunately, we know very little about this emission and nearly nothing about its polarization properties apart from the fact that it is partially polarized. The maximum polarization signal consistent with what we know already is subdominant to galactic emission in all but a few places in the sky in a limited range of frequency. A high fidelity map of pure primordial $B$-mode polarization over a substantial area on the sky requires the ability to discriminate and remove the galactic signal. Removal can be done by observing at multiple frequencies and noting that the galactic emission has a different spectrum or color than the CMB. Models of the foregrounds and the process of foreground removal leave reason for optimism about the prospects of making clean maps. However, knowledge of the real properties of the emission await the results of the currently operating Planck satellite and we currently do not really know how well we can remove foregrounds from CMB maps.

The third experimental element is control of instrument systematic uncertainties. The $B$-mode signal constitutes a particular pattern of polarization but the sought-for strength is nine orders of magnitude weaker than the uniform glow of the CMB itself and 3 orders of magnitude weaker than the $E$-mode polarization. A very subtle shift or rotation of a detector, or glint of radiation from out of the field could mimic a signal. The design of the instruments, the program of mapping and observation, the experimental characterization and calibration of the experiment will all need to be carried out with an unparalleled level of precision.

The current experimental progress in the sensitivity and power of measurement of CMB properties is the key factor that will govern how we carry out the next generation of experiments. Many current experiments are nearly limited by systematic effects which were not anticipated in advance. However, the experiments are done and now we know what must be done to make the next generation instruments even more sensitive. This iterative method is permitting rapid progress on all three elements limiting power of the measurement.

The technology to build sufficiently sensitive experiments is in hand.  Previous gains in sensitivity were the result of new detectors. Further improvement will come from increasing the number of detectors. Currently experiments have ~1000. The next generation will have 10,000 and a proposed satellite has 50,000. The construction of focal planes of this scale and sensitivity will require the sustained support of detector design, development and testing \cite{cmbpol}.

The Planck Satellite \cite{planck}, current operating, will provide new data on polarization within the next 3-4 years, and might even provide the first direct observation of $B$-modes.  Either way, the results from Planck will govern the ultimate design of the next generation of dedicated CMB polarization experiments in space.  From design to construction and launch and operation could take the better part of a decade.  As we enter the second decade of the 21st century, we are thus poised to enter a new realm of precision cosmology---one which could provide a dramatic new window on the early universe and the physical processes that governed its origin and evolution.

\bibliography{cmbpolfinal3}

\bibliographystyle{Science}


\begin{scilastnote}
\item We acknowledge a host of theoretical and experimental collaborators who have helped focus their energy and interest on new ideas that will play a crucial role in the detection of gravitational waves and whose work we have tried to outline here.  Our own research is supported by the Department of Energy Office of Science (Arizona), and NASA, and the NSF for research at Chicago.
\end{scilastnote}


\clearpage

\begin{figure}[]
\centering
\includegraphics[width=.55\textwidth]{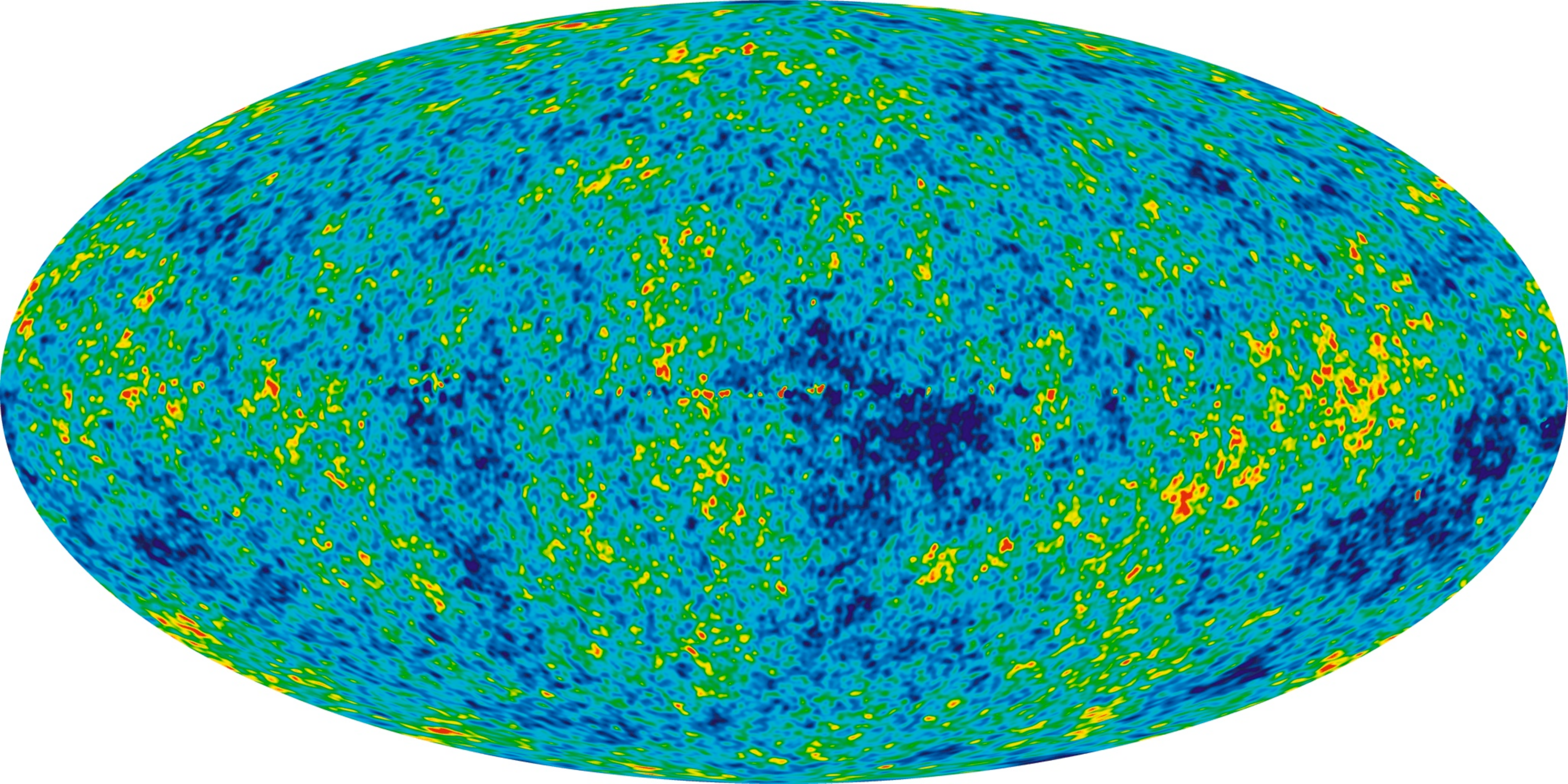}
  \caption{ WMAP 7 Year Map of CMB Temperature Anisotropies across the sky, with galactic foregrounds removed. Color differences represent relative temperature anisotropies on the order of $10^-5$ of the uniform background temperature. Courtesy of WMAP Science Team }
\end{figure}

\begin{figure}[]
\centering
\includegraphics[width=.55\textwidth]{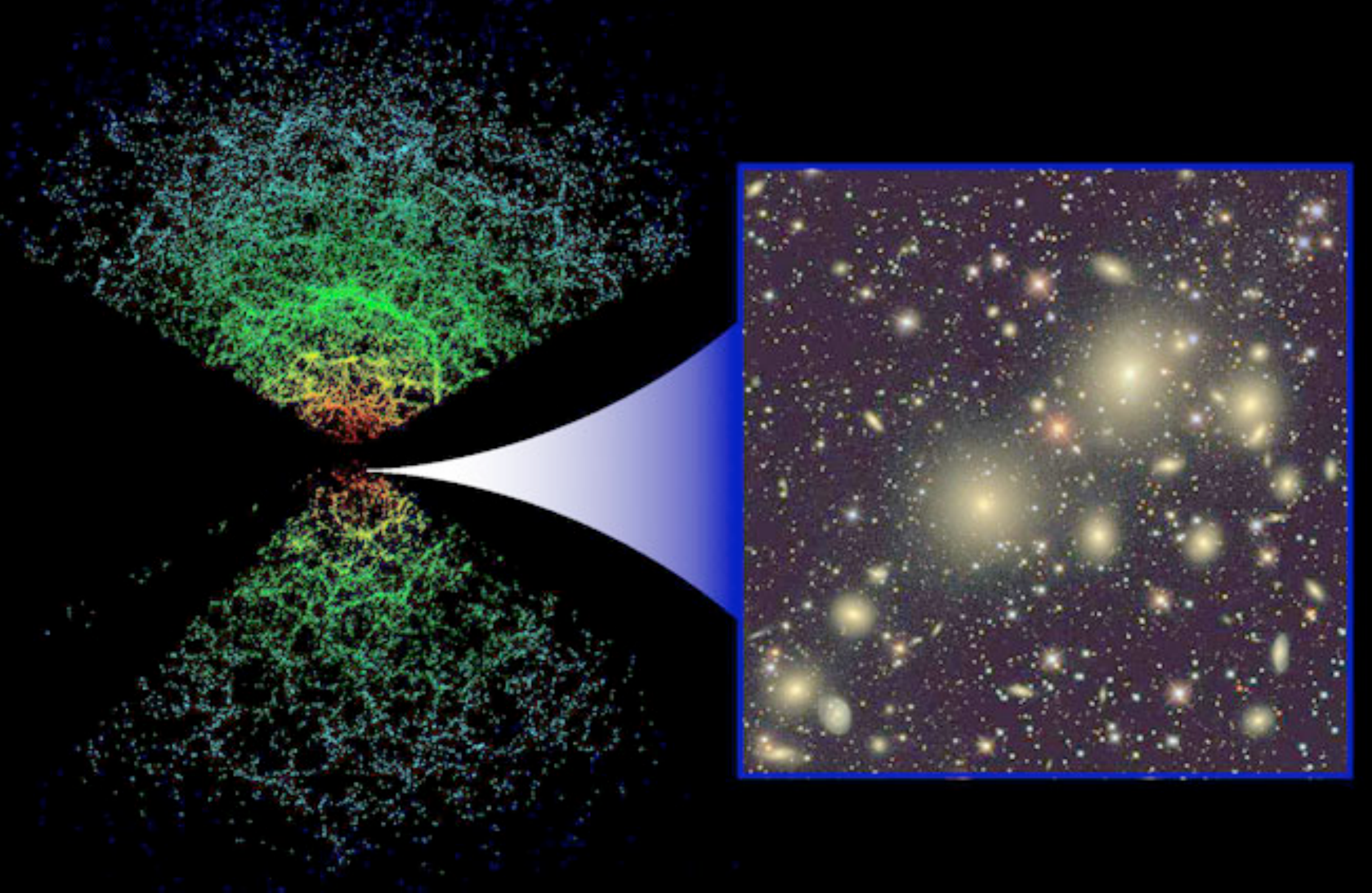}
  \caption{ A map of the observed distribution of galaxies on large scales (from SDSS), illustrating vast structures separated by large voids. The standard cosmological model, built on inflation explains the origin of the WMAP image in Figure 1, and how it evolves into present day distribution of large scale structure. }
\end{figure}

\begin{figure}[]
\centering
\includegraphics[width=.55\textwidth]{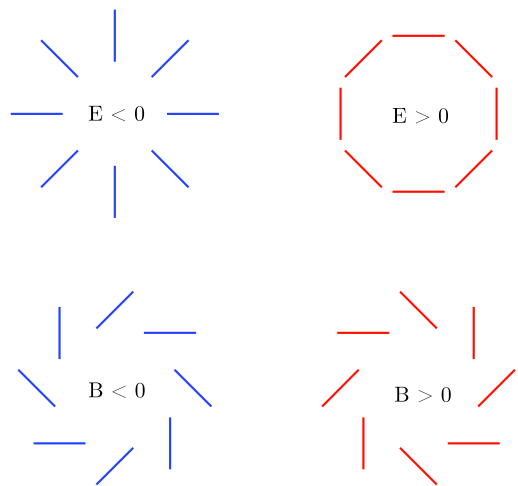}
  \caption{Polarization can be decomposed into $E$- and $B$-modes. The former are radial or tangential with no preferred handedness, akin
to an electric field. Like magnetic fields, $B$-modes do have handedness; Note that if reflected across a line going through the center the $E$-patterns are unchanged, while the positive and negative $B$-patterns get interchanged. This peculiar pattern can be produced only by gravitational waves, not by ordinary density perturbations. }
\end{figure}

\begin{figure}[]
\centering
        \includegraphics [width=.75\textwidth]{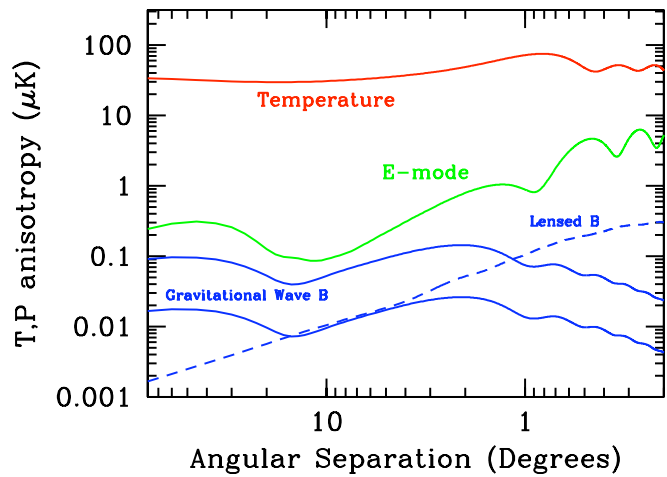}
  \caption{RMS fluctuations in temperature and polarization  of the CMB predicted by inflation. The temperature spectrum has been measured with exquisite precision by dozens of experiments culminating in WMAP, while there have been half a dozen detections of $E$-modes. There are currently only upper limits on the amplitude of $B$-modes.  The upper $B$ mode curve represents the current upper limit $r=0.3$ and the lower curve the value $r=0.01$.}
   \label{fig:anisot}
\end{figure}

\end{document}